\documentclass[preprint,12pt]{elsarticle}

\usepackage{amssymb}

\usepackage[T1]{fontenc}
\usepackage{listings}
\usepackage{hyperref}
\hypersetup{
colorlinks=true,
urlcolor=blue,}

\journal{}

\begin{document}

\begin{frontmatter}


\title{A Brief Note on Building Augmented Reality Models for Scientific Visualization}

\author[inst1]{Mrudang Mathur}
\author[inst2]{Josef M. Brozovich}
\author[inst2,inst3,inst4]{Manuel K. Rausch}

\affiliation[inst1]{organization={University of Texas at Austin, Department of Mechanical Engineering},
            addressline={204 E Dean Keeton Street}, 
            city={Austin},
            postcode={78712}, 
            state={TX},
            country={United States of America}}

\affiliation[inst2]{organization={University of Texas at Austin, Department of Aerospace Engineering and Engineering Mechanics},
            addressline={2617 Wichita Street}, 
            city={Austin},
            postcode={78712}, 
            state={TX},
            country={United States of America}}

\affiliation[inst3]{organization={University of Texas at Austin, Department of Biomedical Engineering},
            addressline={107 W Dean Keeton Street}, 
            city={Austin},
            postcode={78712}, 
            state={TX},
            country={United States of America}}

\affiliation[inst4]{organization={University of Texas at Austin, Oden Institute for Computational Engineering and Sciences},
            addressline={201 E 24th Street}, 
            city={Austin},
            postcode={78712}, 
            state={TX},
            country={United States of America}}

\begin{abstract}


Augmented reality (AR) has revolutionized the video game industry by providing interactive, three-dimensional visualization. Interestingly, AR technology has only been sparsely used in scientific visualization. This is, at least in part, due to the significant technical challenges previously associated with creating and accessing such models. To ease access to AR for the scientific community, we introduce a novel visualization pipeline with which they can create and render AR models. We demonstrate our pipeline by means of finite element results, but note that our pipeline is generally applicable to data that may be represented through meshed surfaces. Specifically, we use two open-source software packages, ParaView and Blender. The models are then rendered through the <model-viewer> platform, which we access through Android and iOS smartphones. To demonstrate our pipeline, we build AR models from static and time-series results of finite element simulations discretized with continuum, shell, and beam elements. Moreover, we openly provide python scripts to automate this process. Thus, others may use our framework to create and render AR models for their own research and teaching activities. 

\end{abstract}

\begin{graphicalabstract}
    \includegraphics[width=\linewidth]{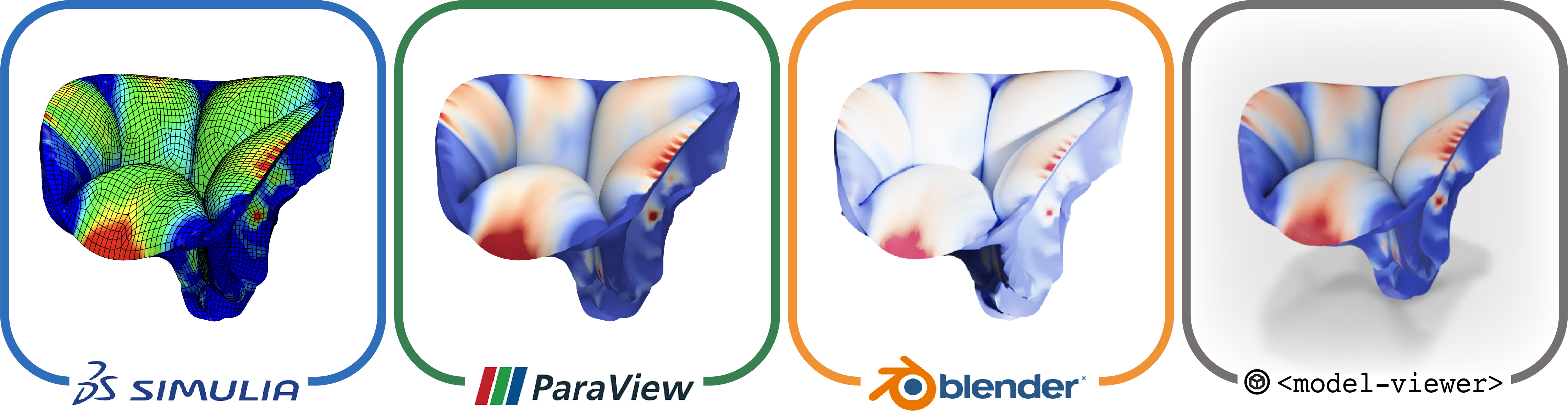}
\end{graphicalabstract}

\begin{highlights}
\item We introduce an open-source pipeline to build and render augmented reality (AR) models for scientific visualization. 
\item We showcase our pipeline by building static and dynamic AR models by means of finite element simulation results.
\item We simplify and automate the creation of AR models through python scripts in Blender.
\end{highlights}

\begin{keyword}
mixed reality \sep virtual reality \sep digital twin \sep metaverse \sep finite elements
\end{keyword}

\end{frontmatter}


\section{Introduction}

Scientific data are often inherently three-dimensional. Examples include imaging results or outputs from numerical methods. However, despite data's inherent three-dimensionality our standard visualization techniques remain two-dimensional, as is the case with figures or videos. Additionally, current visualization techniques lack interactivity. In contrast, Augmented Reality (AR) models can represent the complete spatial and temporal aspects of data, are interactive in nature, and are easily accessible through smartphones. That is, AR is a next-generation visualization technique that overlays computer graphics directly into the physical space surrounding the user, thereby creating an immersive experience\cite{Carmigniani2011}. While originally championed by the entertainment, gaming, and computer graphics communities, AR now finds several applications in manufacturing\cite{Nee2012}, medicine\cite{Chu2012}, and education\cite{Plunkett2019}. 

Despite the clear benefits of AR models, their adoption in research and education has been limited\cite{Huang2015,Hedenqvist2021}. This is, at least in part, due to the use of proprietary software and hardware previously needed to create and render AR experiences, respectively\cite{Sutherland2013}. Additionally, creating AR models often requires specialized training in computer graphics and 3D modeling that is not germane to most disciplines\cite{Huang2017}. Thus, the objective of our current work is to help the scientific community to overcome some of these challenges associated with AR visualization. To do so, we introduce a novel open-source pipeline to build and render AR models of mesh-based scientific data. For demonstration purposes we focus specifically on visualization of finite element results but note that our work is equally applicable to many other disciplines and data types. 


\begin{figure*}
    \centering
    \includegraphics[width=\linewidth]{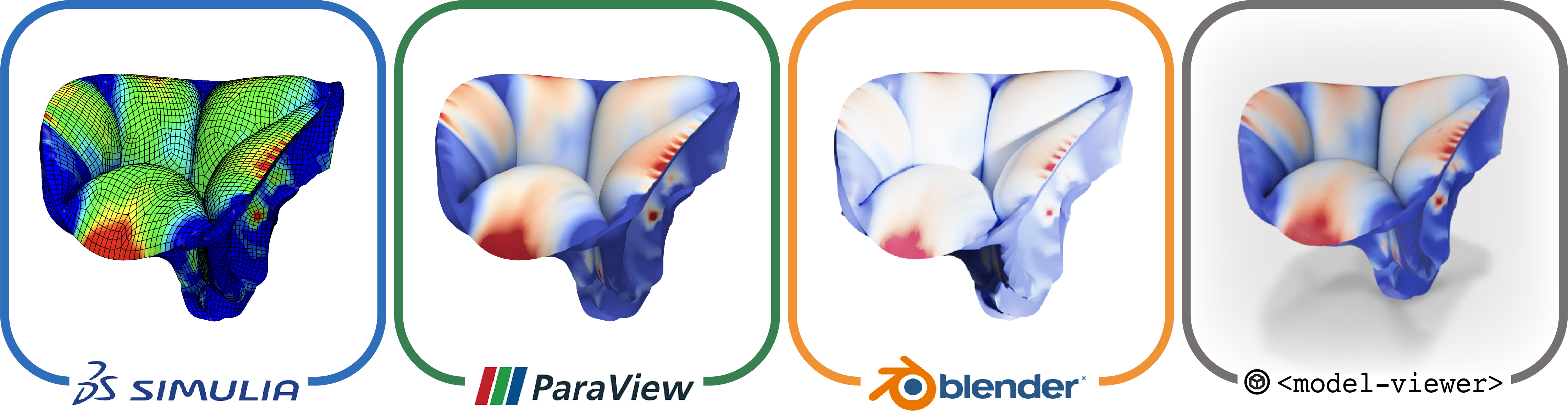}
    \caption{\textbf{Tricuspid valve finite element model rendered through novel AR visualization pipeline}: We transfer simulation results from a finite element solver, Abaqus in this case, to ParaView, an open-source visualization package. In ParaView, we then specify engineering metrics to display and adjust the visualization colormap. Next, we export a surface geometry of our results to Blender, a popular open-source 3D modeling software. In Blender, we calibrate geometry scaling, translation, rotation, as well as lighting for AR visualization and animate our time-series results. Finally, we export AR models that are rendered through <model-viewer> on a smartphone.}
    \label{fig:fig_summary}
\end{figure*}

\section{Material and methods}

Towards building and rendering AR models of mesh-based data, we propose an open-source pipeline using ParaView (Kitware Inc, Clifton Park, NY), Blender (Blender Foundation BV, Amsterdam, The Netherlands), and <model-viewer> (Google Inc, Mountain View, CA), see \textbf{Figure \ref{fig:fig_summary}}. While we focus on finite element analyses in this article, our pipeline can also be used to create AR models of data from non-numerical experiments such as imaging\cite{Abderezaei2021,MOERMAN20091150}. Briefly, we first export our simulation results from a finite element solver, for example Abaqus (v6.20-1, Dassault Systémes, Vélizy-Villacoublay, France), to a file format compatible with ParaView. In ParaView, we then create and export a surface-only version of our simulation mesh and data. Next, we transfer these surface meshes to Blender where we calibrate model size, location, as well as orientation and create animations in case of time-series data. From Blender, we then export Android and iOS-device compatible AR assets that we host on our website using <model-viewer>. Finally, as a demonstration we build AR models containing element types commonly used in computational mechanics: continuum elements and structural elements such as shells and beams. A detailed description of the post-processing steps is provided in the following sections. Additionally, further details on software versions and file formats used are provided in \textbf{\ref{sec:appendix_software}} and \textbf{\ref{sec:appendix_fileformats}}, respectively.

\subsection{Creating a surface mesh}
We first assemble our mesh-based results in the \textbf{*.vtk} file format. Creating \textbf{*.vtk} files is extensively documented in the literature and we direct readers to an excellent guide detailing this process \cite{schroeder2006visualization}. Next, we import the \textbf{*.vtk} files in ParaView and calibrate the visualization to better represent engineering metrics such as stress, strain, and displacement. To this end, we filter our data and apply colormaps to our simulation mesh. We then generate a surface-only mesh of our results. Finally, we export the surface mesh of our model as a binary \textbf{*.ply} file. Please note, we embed any engineering metrics as vertex colors in these files. We provide example \textbf{*.ply} files for continuum, shell, and beam elements on the GitHub repository supplemental to this article. Additionally, see \textbf{\ref{sec:appendix_surface}} for a list of filters used to create surface meshes of the aforementioned elements in ParaView.

\begin{figure*}
    \centering
    \includegraphics[width=\linewidth]{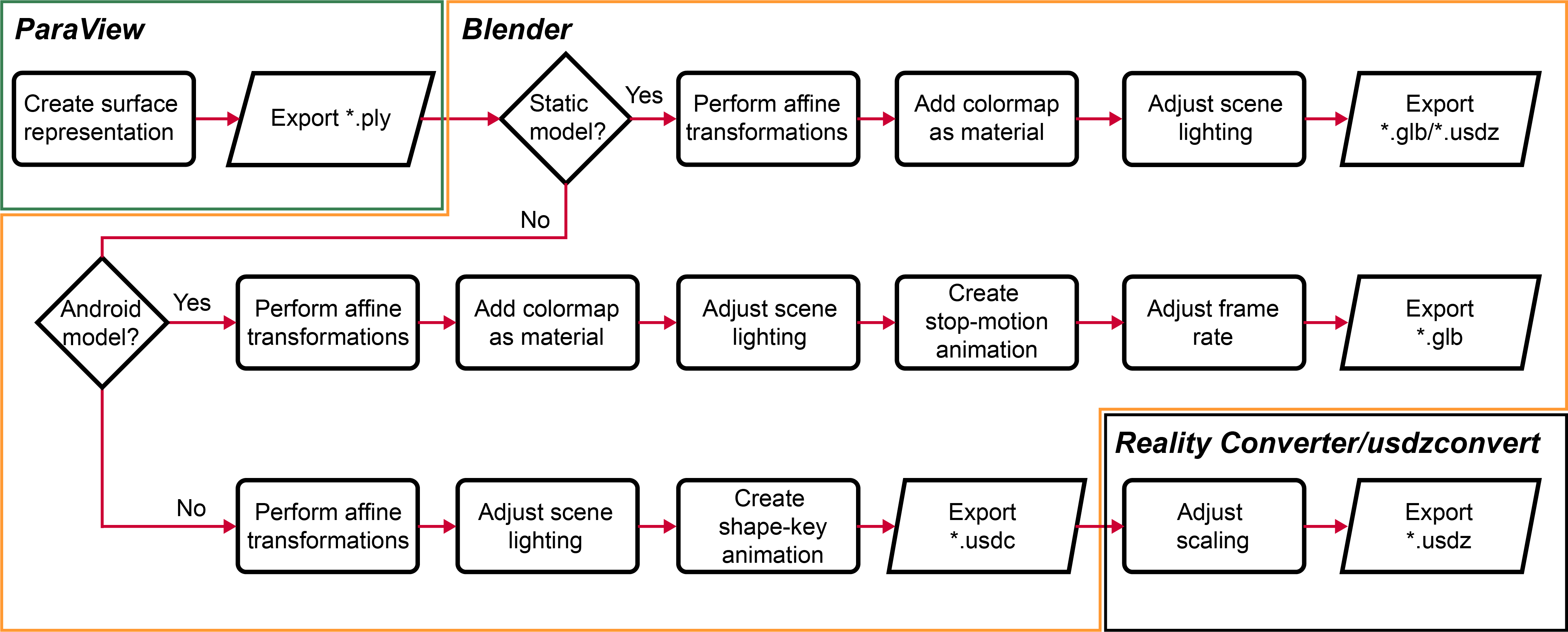}
    \caption{\textbf{Post-processing operations in Blender:} We process surface geometries from ParaView in three distinct ways based on AR model type and rending platform. To this end, we classify AR models as static models, dynamic AR models for Android devices, and dynamic AR models for iOS devices. We customize model size, position, and lighting in all cases, adjust animation timing for time series data, and add vertex colors to all models except dynamic models for iOS (see our discussion section for more detail for why we omit colors in iOS dynamic models).}
    \label{fig:fig_flowchart}
\end{figure*}

\subsection{Building AR models}

To create AR models from surface meshes of our finite element results, we follow a multi-step process in Blender, as described in \textbf{Figure \ref{fig:fig_flowchart}}. Here, we employ a common process to build static AR models for all mobile platforms. However, because Android and iOS devices use ARCore and ARKit as rendering libraries, respectively, they require distinct treatment when visualizing time-varying data. Thus, our process differs when animating our time-series data results in dynamic AR models for Android and iOS devices.

\subsubsection{Static models}

To build a static AR model, we first import the \textbf{*.ply} file in Blender where we adjust model scaling, translation, and rotation. To visualize any colormaps that we add to our model in ParaView, we first create a "Material" within Blender. "Materials" are data structures in Blender that store and govern the appearance and texture of models. To our "Material", we assign vertex colors that are embedded in the \textbf{*.ply} file. Next, we apply a "Decimate Modifier" to our geometry. This reduces the number of polygons in our AR assets which, in turn, reduces their storage size. Furthermore, we apply a "Solidify" modifier to shell geometries. This enhances model visibility while rendering.  Additionally, we adjust lighting around our model in 3D space. Finally, we export an Android-compatible \textbf{*.glb} file through Blender's native exporter. To generate an iOS-compatible \textbf{*.usdz} file we use the BlenderUSDZ add-on via the Blender GUI. To simplify the model creation process, we have also automated the above described steps using Python. See \textbf{Supplementary Scripts 1a-c} to create static AR models of simulations with continuum, shell, and beam elements, respectively.

\subsubsection{Dynamic models: Android}
First, we follow the same steps used to create static models. That is, we import all geometries of our time-series data, perform affine transformations, and add the appropriate colormaps to our models in Blender. Next, we create a stop-motion animation of our time-series results. To this end, we ensure that only a single geometry is visible in each frame of our animation. Thereby, we emulate mesh motion between each animation frame. To achieve this, we dynamically resize each geometry in our dataset over the time-span of the animation. Finally, we adjust model lighting and exported a \textbf{*.glb} file of our animation for rendering on Android devices. We provide \textbf{Supplementary Scripts 2a-c} to automatically build dynamic AR models of the continuum, shell, and beam element examples.

\subsubsection{Dynamic models: iOS}
To build dynamic models for iOS devices, we first imported all \textbf{*.ply} files into Blender. Next, we adjusted the scaling, translation, and rotation of our models. We then animated our time-series data using Shape Keys. To this end, Blender automatically builds a material point correspondence between each mesh in our dataset. As a result, we exactly animate the motion of each vertex in our model geometry over time. Please note, Shape Keys require the number of polygons remains constant between meshes. Thus, special attention should be paid to simulation results with self-contacting geometries to ensure a constant polygon count. Next, we modified scene lighting and exported an intermediate \textbf{*.usdc} file of our animated model. Finally, we adjusted model scaling and converted our animated model to \textbf{*.usdz} files in Reality Converter or \emph{usdzconvert} in MacOS and Windows/Linux systems, respectively. See \textbf{Supplementary Scripts 3a-c} to automate Shape Key animations in Blender. 

\subsection{Hosting and rendering models}
We use GitHub to host, access, and share our AR models. To this end, we integrate <model-viewer> with our research webpage to render AR assets. We chose <model-viewer> over other, similar, rendering platforms as it is free to use, compatible with both Android and iOS devices, and can be accessed without app download. Finally, we share links to our AR models through QR codes. This choice is motivated by the relative ease of generating QR codes, accessing them via smartphone, and by the possibility of embedding QR codes in presentations, videos, and figures.  

\section{Results}

\begin{figure*}[ht!]
    \centering
    \includegraphics[width=\linewidth]{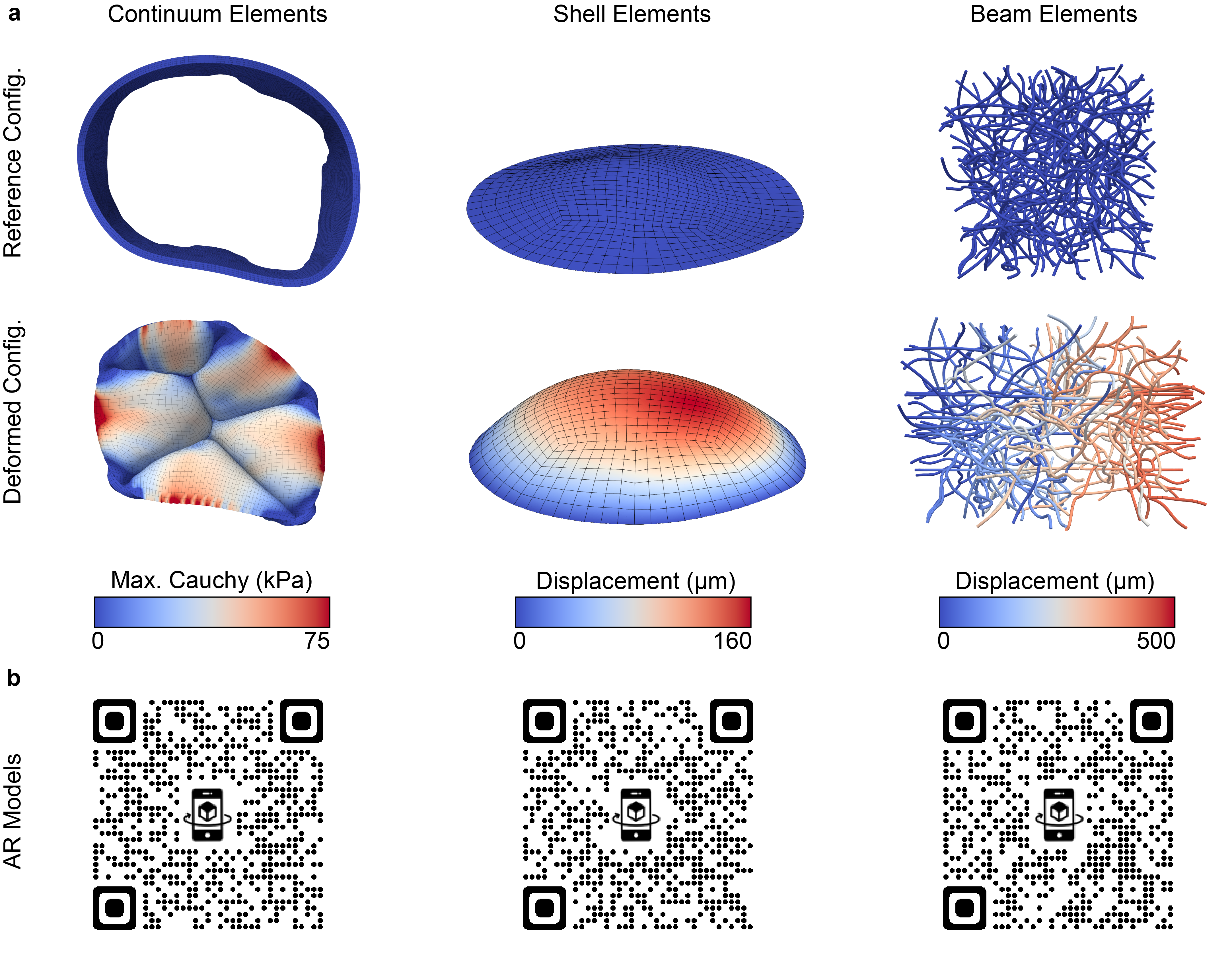}
    \caption{\textbf{AR models of finite element simulations discretized with continuum, shell, and beam elements:}(a) We present simulation results in the reference and deformed configurations which are overlaid with contours of engineering metrics, such as maximum principal Cauchy stress and displacement magnitude. (b) Static and dynamic models of each simulation can be accessed through the associated QR code.}
    \label{fig:fig_results}
\end{figure*}

\begin{figure*}[ht!]
    \centering
    \includegraphics[width=\linewidth]{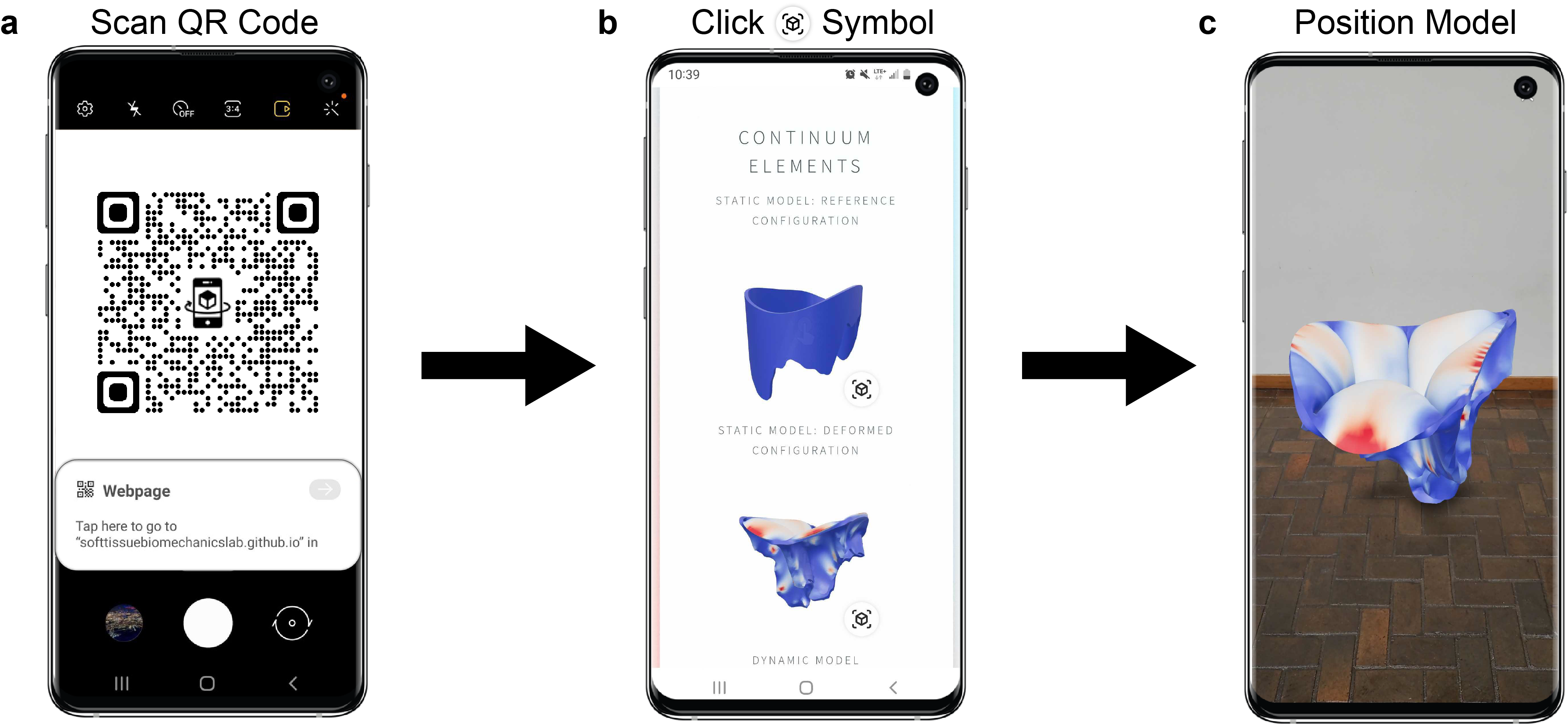}
    \caption{\textbf{Accessing AR models:} To view AR models on a smartphone (a) we first scan the QR code and load the associated webpage, (b) we then click the <model-viewer> logo to enable AR, and finally (c) we position the AR model in 3D space around us.}
    \label{fig:figure_AR}
\end{figure*}

To demonstrate our open-source visualization pipeline, we built AR models of three distinct finite element analyses from our research group, see \textbf{Figure \ref{fig:fig_results}a}. These include a dynamic model of the human tricuspid valve\cite{Mathur2022}, a bulge inflation test of tissue from a rat's anterior tricuspid valve leaflet\cite{Meador2022}, and the uniaxial extension of an undulated fiber network. These simulations are discretized with continuum, shell, and beam elements, respectively. Here, we present images of the reference and deformed configuration of each simulation, as one would see in a scientific figure. Additionally in \textbf{Figure \ref{fig:fig_results}b}, we provide QR codes leading to a webpage containing static and dynamic AR models of each simulation. Readers are encouraged to scan these QR codes and view the linked AR models as described by \textbf{Figure \ref{fig:figure_AR}}. Using our open-source pipeline, we successfully build and render AR models of the aforementioned finite element analyses. Thereby, we provide scientists, in general, and mechanicians, in particular, an avenue to create and share their results in all spatial and temporal dimensions.

\section{Discussion}
Augmented reality represents the next frontier in personal computing and has the capacity to transform scientific visualization. In this brief note, we introduced an open-source visualization pipeline to build and render AR models from finite element simulation results. Our pipeline integrates an established and versatile tool in scientific visualization (Paraview) with a popular 3D modeling tool (Blender) to standardize and simplify the process of building AR models from scientific data. We then use an open-source AR platform (<model-viewer>) to access and render our models on Android and iOS devices. We also demonstrated our pipeline on results from finite element simulations spatially discretized with continuum, shell, and beam elements. Furthermore, we provide python scripts to automate the creation of those same models. Thus, we consider this pipeline to be successful in achieving our stated objectives: to democratize and simplify AR visualization of three-dimensional data by means of finite element results. That is, we eschewed the need for proprietary software, such as NVIDIA Omniverse, and expensive AR headsets, such as the Microsoft Hololens. Moreover, through the provided python scripts, we significantly reduce the training required to use an open-source, industry-standard 3D modeling tool to create AR assets.   

In addition to fulfilling these objectives, our pipeline can serve as a cornerstone for future studies integrating scientific data with mixed-, virtual-, and augmented-reality applications. Specifically concerning results from mechanical analyses we'd like to note that our methods can be easily extended to Eulerian grids and do not have to be limited to Lagrangian analyses as used in our examples\cite{Shad2021,Pfaller2021,Menon2019,BERGER2022103742}. Furthermore, we can build AR models from isogeometric analysis results due to Blender's in-built support for NURBS\cite{Dortdivanlioglu2018,SHEPHERD2022113602,CHEMIN201577}. Additionally, we may interface Blender with packages such as Unity to design virtual reality experiences\cite{Bhatia2020}. Finally, when combined with image processing packages, our methods can be used to effectively visualize and interact with digital twins\cite{Revetria2019,Febrianto2021}.   

Naturally, our methods are subject to certain limitations. Firstly, our current method of creating dynamic \textbf{*.usdz} files precludes the use of custom colormaps. This is currently a limitation of Apple's underlying USD codebase. We anticipate that dynamic vertex colors will be added in upcoming software releases. Secondly, our current \textbf{*.usdz} animation technique requires users to specify the number of times an animation should repeat. This increases the file size of the AR asset, thereby making it harder to access over slower internet connections. Moving forward, we aim to programmatically loop \textbf{*.usdz} animations, similar to those in \textbf{*.glb} files. Finally, by using <model-viewer> to render AR models, we require users to create both \textbf{*glb} and \textbf{*.usdz} models of any scientific results they have. In the future, we anticipate a greater cross-compatibility in files between Android and iOS systems, thereby reducing this burden on content creators.


\section{Conclusion}

In conclusion, we introduced an end-to-end pipeline for building and rendering AR models from scientific, mesh-based data\cite{Portela2018}. Importantly, our pipeline only uses open-source software packages and, thus, allows users to freely access AR models on any smartphone through the internet. Furthermore, we showcased our pipeline by building AR models of finite element analyses with three common element discretizations in used computational mechanics. Moreover, we provided python scripts to automatically create those models. Through this work, we hope to simplify and accelerate the adoption of AR visualization in scientific visualization. Thereby enabling researchers, educators, and students to gain a deeper understanding of complex spatio-temporal results associated with their data. Importantly, all scripts and information necessary to reproduce our work are openly available through a GitHub repository listed under ``Code Availability'' below.   


\section*{Acknowledgements}
We appreciate support from the American Heart Association through an award to Dr. Rausch (18CDA34120028) and a predoctoral fellowship to Mrudang Mathur (902502), as well as the National Institutes of Health through an award to Dr. Rausch (1R21HL161832). We would also like to thank Soham M. Mane for sharing the results of his fiber network simulations with us. 

\section*{Disclosures}
Manuel K. Rausch has a speaking agreement with Edwards Lifesciences. None of the other authors have conflicts of interest to disclose.

\section*{Author Contributions}
MKR and MM wrote the manuscript. MM and JMB developed the code. MM produced all figures and tutorials. All authors reviewed the manuscript.  

\section*{Code Availability}
All supplementary scripts, surface geometries, and AR model examples are available in the GitHub repository associated with this article. \\ URL: \url{https://github.com/SoftTissueBiomechanicsLab/AR_Pipeline.git}
\appendix

\section{Software requirements}
\label{sec:appendix_software}
Our proposed pipeline requires the use of multiple open-source software and is, thus, subject to compatibility errors across different software versions. Therefore, we have compiled a list of stable and compatible software packages. These packages, along with their download links, are detailed in the GitHub repository associated with this article. The packages are: 

\begin{itemize}
    \item ParaView 5.10
    \item Blender 2.83
    \item BlenderUSDZ add-on
    \item Apple Reality Converter (for MacOS users)
    \item Apple usdzconvert utility (for Windows/Unix users)
\end{itemize}

\section{File formats}
\label{sec:appendix_fileformats}
To ensure the flexibility of our methods to varied finite element solvers, visualization packages, and mixed-reality platforms we use several industry-standard and open-source file formats in our pipeline. Specifically, we employ \textbf{*.vtk} files for storing and accessing numerical simulation data and \textbf{*.ply} files to store model geometries. Moreover, we render our AR assets as \textbf{*.glb} and \textbf{*.usdz} files for Android and iOS devices, respectively. 

\section{Generating surface meshes in ParaView}
\label{sec:appendix_surface}
To generate surface meshes for results with continuum and shell elements we use the following filters:
\begin{verbatim}
1. Filters > Alphabetical > Extract Surface
2. Filters > Alphabetical > Generate Surface Normals
\end{verbatim}
For beam elements, we first employ the Tube filter in ParaView viz.:
\begin{verbatim}
1. Filters > Alphabetical > Tube
2. Filters > Alphabetical > Extract Surface
3. Filters > Alphabetical > Generate Surface Normals
\end{verbatim}

 \bibliographystyle{elsarticle-num} 
 \bibliography{References}

\end{document}